\documentclass[reprint,amsmath,amssymb,aps,pre,floatfix]{revtex4-2}
\usepackage[colorlinks=true, allcolors=blue]{hyperref}
\usepackage{graphicx}
\usepackage{dcolumn}
\usepackage{bm}
\usepackage{array}
\usepackage{ulem}
\usepackage[dvipsnames]{xcolor}

\usepackage[capitalise]{cleveref}
\usepackage{upgreek}

\newcommand{\Wcmsqd}{{\mathrm{W }\mathrm{cm}^{-2}}}
\usepackage{lipsum}

\begin{document}
\title{Impact of electron trapping on stimulated Raman scattering under incoherent broadband laser light in homogeneous plasma}
\author{David R. Blackman$^{1}$, Vladimir Tikhonchuk$^{1,2}$, Ondrej Klimo$^{1,3}$, Stefan Weber$^{1}$}
\affiliation{$^{1}$ELI ERIC, ELI Beamlines, Dolni Brezany, Czech Republic}
\affiliation{$^{2}$ Centre Lasers Intenses et Applications, Université Bordeaux 1-CNRS-CEA, 33405 Talence Cedex, France}
\affiliation{$^{3}$ Faculty of Nuclear Sciences and Physical Engineering, Czech Technical University in Prague, Brehova 7, 115 19 Praha 1, Czech Republic}

\preprint{APS/123-QED}
\date{\today}
\begin{abstract}
Backward stimulated Raman scattering is a three-wave coupling instability requiring the matching of an incoming pump light wave to a scattered light wave and electron plasma wave. It can be harmful to laser-driven inertial confinement fusion because of the reflection of a part of incident laser flux and the generation of suprathermal electrons that preheat the fuel. It is believed that by increasing the laser bandwidth one can suppress the excitation of Raman scattering and mitigate its detrimental effects. It is demonstrated in this paper that using a broad bandwidth laser has little effect on stimulated Raman scattering in the kinetic inflation regime where Landau damping dominates, as the additional bandwidth allows the electron plasma wave to match a wider range of laser frequencies. As a result, plasma wave saturation and Raman backscattering levels remain high even when the laser bandwidth is much larger than the effective instability growth rate.
\end{abstract}
\maketitle
\section{Introduction}\label{sec1}
The use of coherent, broad bandwidth light has been suggested to mitigate the growth of parametric instabilities~\cite{Thomson1974, Thomson1975, Obenschain1976} as far back as the mid-70s. Several variants of the pump bandwidth have been considered: a pump wave containing several slightly differing frequencies~\cite{Milovich1984, Zhao2017}, smoothing by spectral dispersion \cite{Skupsky1989}, where the phase-modulated light pulse is dispersed spatially, a low coherent light~\cite{Santos2007, Brandao2008}, where the wave phase and amplitude varying in time, or STUD pulses~\cite{Albright2014} where laser light is unevenly modulated temporally and spatially. 

Motivated by the need for mitigation of parametric instabilities in laser-driven inertial confinement fusion (ICF), low-coherence laser light has been studied intensely for the last two decades, by teams in the US~\cite{Bates2018, Follett2018, Follett2019, Seaton2022}, China~\cite{Mufei2022,Zhao2022, Guo2023, Zhao2024}, and Europe~\cite{Fusaro2021, Brandao2021}. These intense studies provided the basis for the development of experimental facilities able to produce light with bandwidth and intensity relevant to ICF~\cite{Eimerl2014, Dorrer2022, Gao2020, Gao2020a}. However, the first experiments with a broadband laser are producing a variety of results~\cite{Wang2024, Lei2024}, which are not yet fully understood.

While the suppression of ion-based instabilities, stimulated Brillouin scattering (SBS) and filamentation, by small bandwidths is apparent~\cite{Bates2018, Follett2018, Follett2019, Seaton2022}, the suppression of electron-based instabilities, stimulated Raman scattering (SRS) and two-plasmon decay, is less certain. In some scenarios, the amplitude spikes due to laser incoherence can be seen to cause hot electron bursts seeding later SRS \cite{Liu2022}, or switch backward scattered light to side-scatter~\cite{Li2023}. Additionally, by suppressing SBS, SRS is allowed to grow further due to the decreased laser reflectivity~\cite{Liu2023}. Early work on broadband laser-plasma interactions in inhomogeneous plasmas~\cite{Guzdar1991} identified a scenario where the length of SRS resonant amplification in a plasma increases due to an increase in the laser bandwidth, offsetting any reduction in the growth rate by an increase in resonance.

In this paper, we identify another factor affecting the SRS response on the laser bandwidth, that of the plasma wave spectral modification due to trapped electrons. The dispersion of the electron plasma wave, mediating the coupling of the pump and the scattered wave, is important when considering SRS response in plasma. This is of particular importance in the inflation regime~\cite{Fernandez2000, Yin2006, Riconda2011, Spencer2020}, where the Landau damping of plasma wave dominates at the initial stage of the growth. However, the subsequent modification of the plasma wave frequency and damping due to the electron trapping results in the resonance detuning and a flashy behavior where the SRS reflectivity switches on and off. A study of the SRS excitation in a low-density inhomogeneous plasma in the kinetic inflation regime~\cite{Mufei2022} shows that there is a synergy between the laser bandwidth and the variation in plasma frequency due to a density gradient such that a wider range of matching frequencies is allowed without a significant reduction in SRS. 

The study presented here demonstrates that kinetic effects alone are enough to reduce the effectiveness of broadband on suppressing SRS, and that both the plasma wave frequency alteration and the bandwidth of the plasma wave itself, likely play a role in the three wave coupling.

Section \ref{sec2} recalls the basic elements of the linear theory of SRS driven by a monochromatic or broadband pump wave needed for further analysis. Results of numerical simulations with a particle-in-cell code and their analysis are presented in Sec. \ref{sec3}, while Sec. \ref{sec4} contains a general discussion and conclusions. 

\section{Linear theory of SRS}\label{sec2}
\subsection{Monochromatic light}\label{sec21}
The general analysis of parametric instabilities can be found in Ref. \onlinecite{Kruer2003, Tikhonchuk2024}. For the backward SRS, the resonant wavenumber for a plasma wave $k_p$ driven by a monochromatic electromagnetic wave with a frequency $\omega_0$ can be written as:
\begin{equation}\label{eq:kback}
k_p = \frac{\omega_0}c\,\left( \sqrt{1-\frac{\omega_\mathrm{pe}^2}{\omega_0^2}}+\sqrt{1-2\frac{\omega_\mathrm{pe}}{\omega_0}}\right),
\end{equation}
where $c$ is the light velocity, $\omega_\mathrm{pe}$ is the plasma frequency, and $k_0=(\omega_0^2-\omega_{\rm pe}^2)^{1/2}/c$ is the laser wavenumber. Neglecting dissipation, the growth rate of SRS $\gamma_\mathrm{SRS}$ is given by the following equation:
\begin{equation}\label{eq:gsrs}
\gamma_\mathrm{SRS} = \frac {eE_0 k_p}{4m_e \omega_0} \sqrt{\frac{ \omega_\mathrm{pe}}{\omega_0-\omega_\mathrm{pe}}},
\end{equation}
where $e$ and $m_e$ are the electron charge and mass, and $E_0$ is the laser field amplitude. This is an idealized expression, which assumes the plasma parameters are invariant in space and time. 

An expression for the SRS growth rate, which takes into account the damping of plasma and electromagnetic waves, can be found as a solution to the following equation:
\begin{equation}\label{eq:gdamp}
(\gamma +\gamma_\mathrm{pe} )(\gamma + \gamma_\mathrm{em}) =\gamma_\mathrm{SRS}^2,
\end{equation}
where $\gamma_\mathrm{pe}$ is the damping rate of plasma wave, including the collisional and Landau damping, and $\gamma_\mathrm{em}$ is the collisional damping of scattered electromagnetic wave. For the conditions we consider in this paper, which correspond to laser intensities around $I_l\sim10^{15}$~W cm$^{-2}$ and plasma densities around 5\% the critical density, collisional damping can be neglected, while Landau damping of plasma waves can overcome the growth rate of SRS leading to a reduction of the SRS growth rate. However, according to Eq. \eqref{eq:gdamp}, SRS is not completely suppressed, and steadily growing plasma waves may eventually modify the electron distribution function, suppress Landau damping, and accelerate the instability growth rate. This is called the inflation regime of SRS~\cite{Vu2002, Yin2006, Riconda2011}. However, electron trapping is also associated with a reduction of plasma wave frequency, detuning the resonance and SRS quenching. Consequently, the inflation regime results in a periodic emission of scattered waves in the form of flashes. The growth of SBS may compete by increasing the intensity of laser light and reducing the SRS growth rate.

\subsection{Dispersion equation for SRS driven by broadband laser light}\label{sec22}
The SRS growth can also be reduced by the pump bandwidth. Here, we consider a broadband laser consisting of $N$ co-propagating spectral components with regular frequency spacing. The spectral components have the same amplitude and polarization with the frequency shift $\omega_b$ with respect to each other. The expression for the laser field propagating along the $x$ axis and polarized in the $z$ direction reads:
\begin{align}\label{eq:mmode}
E_l &=E_0 N^{-1/2} \\ \times {\rm Re}\,\sum_{n=-N/2}^{N/2} \nonumber &\exp[ik_0x-i\omega_0t -i n\omega_b (t-x/v_g) +i\phi_n],
\end{align}
where $E_0$ is the average laser amplitude, $v_g$ is the laser group velocity, and the phases $\phi_n$ take random values in the interval $(0,2\pi)$. They verify the following conditions:
\begin{equation}\label{eq:delta}
 \langle {\rm e}^{i\phi_n}\rangle = 0 \quad {\rm and} \quad \langle {\rm e}^{i\phi_n-i\phi_{n'}}\rangle = \delta_{n,n'},
\end{equation}
where $\delta_{n,n'}$ is the Kronecker symbol, and the angle brackets indicate the average over the phases of all spectral components. Then, the equation for the Stokes Fourier component of the backscattered wave reads:
\begin{align}\label{eq:es}
D_s&(k,\omega) \, E_s(k,\omega) = \frac{\omega_{\rm pe}^2 E_0}{2 n_e N^{1/2}} \sum_n {\rm e}^{i\phi_n} \nonumber \\ \times & \delta n_e^\ast(k_0-k+ nk_b,\omega_0- \omega+n\omega_b) ,
\end{align}
where $D_s(k,\omega) =\omega^2-k^2c^2-\omega_{\rm pe}^2$ is the dispersion relation for the scattered wave, $k_b= \omega_b/v_g$ and $\delta n_e$ is the density perturbation with respect to the background density $n_e$ related to the plasma wave, which is driven by the interference of the incident and scattered electromagnetic waves:
\begin{align}\label{eq:dne}
 D_p&(k,\omega)\, \delta n_e(k,\omega) = \frac{e^2k^2 E_0 n_e}{2m_e^2\omega_0^2 N^{1/2}} \sum_n {\rm e}^{i\phi_n} \nonumber \\ 
 \times& E_s^\ast(k_0-k+nk_b,\omega_0-\omega+ n\omega_b),
\end{align}
where $D_p(k,\omega)=\omega^2-\omega_{\rm ek}^2(k)+2i\omega \gamma_{\rm pe}$ is the dispersion relation for the electron plasma waves, $\omega_{\mathrm{ek}}=\omega_\mathrm{pe} +(3/2)k^2 v_{\rm Te}^2/\omega_{\rm pe}$ is the Bohm-Gross frequency, $v_{\rm Te}= (T_e/m_e)^{1/2}$ is the electron thermal velocity, and $T_e$ is the electron temperature.
Excluding the amplitudes of scattered waves from this system of equations and neglecting higher order terms containing $2k_b$ and $2\omega_b$, we find an equation for the plasma wave amplitudes:
\begin{align}\label{eq:dne2}
D_p&(k,\omega)\,\delta n_e(k,\omega) = \frac{e^2k^2\omega_{\rm pe}^2 E_0^2}{4m_e^2 \omega_0^2N} \sum_{n,n'} {\rm e}^{i\phi_n-i\phi_{n'}} \nonumber \\ 
&\times \frac{ \delta n(k+(n'-n)k_b, \omega+(n'-n)\omega_b)}{D_s(k_0-k+nk_b,\omega_0-\omega+n\omega_b)}.
\end{align}
These equations contain random phases. Performing the phase average according to Eq. \eqref{eq:delta}, the off-diagonal terms are factored out, and the dispersion equation can be simplified to:
\begin{align}\label{eq8}
D_p&(k,\omega)= \frac{e^2k^2\omega_{\rm pe}^2E_0^2 }{4 m_e^2 \omega_0^2N} \nonumber \\ &\times\sum_n  D_s^{-1}(k_0-k+nk_b,\omega_0-\omega+n\omega_b).
\end{align}
This is the generalization of the SRS dispersion equation to the case of a multi-mode pump. It can be simplified assuming the laser bandwidth $N\omega_b$ is much smaller than the electron plasma frequency. Since the frequency of driven plasma wave $\omega$ should be close to the electron plasma frequency, the left-hand side can be simplified as
$$ D_p(k,\omega)\approx 2\omega_{\rm pe} (\delta\omega+i\gamma_{\rm pe}), $$
where $\delta\omega = \omega-\omega_{\rm ek}(k)$ is the detuning from the Bohm-Gross frequency. Similar simplification applies to the dispersion relation of the scattered wave:
$$ D_s(k_0-k,\omega_0-\omega) \approx 2(\omega_0-\omega_{\rm pe})\, [\Delta\omega(k_0-k)-\delta\omega], $$
where $\Delta\omega(k)=\omega_0-\omega_{\rm ek}(k) -\omega_s(k_0-k)$ is the frequency detuning from the resonance and $\omega_s(k)=(c^2k^2+\omega_{\rm pe}^2)^{1/2}$ is the frequency of the scattered wave. Consequently, the dispersion equation \eqref{eq8} reads:
\begin{align}\label{eq8a}
\delta\omega+i\gamma_{\rm pe}=\frac{\gamma_{\rm SRS}^2}{N}\sum_n [\Delta\omega(k)-\delta\omega+\alpha n\omega_b]^{-1},
\end{align}
where $\gamma_{\rm SRS}$ is given by Eq. \eqref{eq:gsrs} and $\alpha=1-v_{gs}/v_g$ accounts for the difference in the group velocities of the pump and scattered waves. In the case of backward SRS,
$$ \alpha =1+(1+k_p/k_0)/(1-\omega_{\rm pe}/\omega_0).$$
For a single spectral component, $N=1$ and $n=0$, Eq. \eqref{eq8a} gives the SRS growth rate considered in Sec. \ref{sec21} at the resonance $\Delta\omega(k_p)=0$, where $k_p$ is given by Eq. \eqref{eq:kback} in the limit $k_pv_{\rm Te} \ll \omega_{\rm pe}$. Assuming $N\gg1$, the sum in this equation can be replaced with an integral, which provides the following expression for the dispersion equation:
\begin{align}\label{eq8c}
 \delta\omega +i\gamma_{\rm pe}= \frac{\gamma_{\rm SRS}^2(k)}{\Omega_{\mathrm{b}}} \ln \frac{\Delta\omega(k)- \delta\omega+ \Omega_{\mathrm{b}}/2}{\Delta\omega (k)- \delta\omega- \Omega_{\mathrm{b}}/2},
\end{align}
where $\Omega_b= \alpha N\omega_b$ is the effective pump bandwidth. 

Similar to the single mode case, the maximum growth rate is achieved at the resonance, $\Delta \omega(k_p)=0$. Introducing $\delta\omega=i\gamma$, Eq. \eqref{eq8c} can be written as
\begin{align}\label{eq8d}
\gamma +\gamma_{\rm pe}= \frac2\Omega_b \gamma_{\rm SRS}^2\, {\rm Arctan}\frac{\Omega_b}{2\gamma}. 
\end{align}

\begin{figure}[!h]
\includegraphics[width=0.4\textwidth]{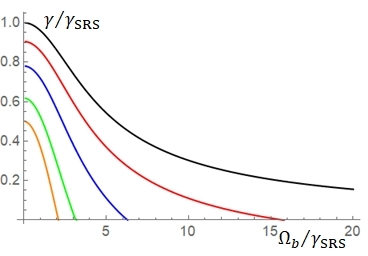}
 \caption{Dependence of the SRS growth rate on the laser pump bandwidth for the Landau damping rate $\gamma_{\rm pe}/\gamma_{\rm SRS}=0$ (black), 0.2 (red), 0.5 (blue), 1.0 (green), and 1.5 (orange).}
  \label{fig0}
\end{figure}
In the limit of small bandwidth, $\Omega_b <\gamma_{\rm SRS}$, this equation can be written as
\begin{align}\label{eq13} 
\gamma (\gamma+\gamma_{\rm pe})= \gamma_{\rm SRS}^2(1-\Omega_b^2/12\gamma^2).
\end{align}
The bandwidth leads to a slight reduction in the growth rate. Conversely, the effect of bandwidth is strong in the opposite limit, $\Omega_b \gg \gamma_{\rm SRS}$, where this equation reads:
$$ \gamma+\gamma_{\rm pe}= \pi \gamma_{\rm SRS}^2/\Omega_b. $$
In particular, SRS is completely suppressed for $\Omega_b \geq \pi \gamma_{\rm SRS}^2/\gamma_{\rm pe}$. The numerical solution to the dispersion equation \eqref{eq8d} is shown in Fig. \ref{fig0}.
However, these predictions for the SRS gain suppression obtained from the linear theory are strongly modified in the inflation regime, where the electron trapping affects the Landau damping and dispersion of the plasma wave.

A similar multi-mode model of excitation of parametric instabilities applies to the forward SRS and backward SBS with appropriate expressions for the resonance matching $\Delta\omega(k)$ and coupling coefficient. 

\section{Numerical Simulations}\label{sec3}
\subsection{Input parameters}\label{sec31}
The kinetic particle-in-cell (PIC) code {\sc Smilei} \cite{Derouillat2018} is used in this study. The plasma and laser parameters common to all simulations are displayed in~\cref{tab:simulation_1}. The cell size is set to the Debye length for an electron temperature of 250~eV and $n_e/n_{\rm cr}=0.05$, where $n_{\rm cr}$ is the electron critical density. So that the electron temperature can be varied significantly whilst maintaining a modest box size in one-dimensional simulations.

The broadband laser field used in this study is described as a sum of multiple frequencies with randomized phases and is simulated at the $x=0$ boundary according to Eq. \eqref{eq:mmode}.
The phase $\phi_b$ is chosen from a uniform random distribution and is fixed across all simulations. A single mode laser field, described by Eq. \eqref{eq:mmode} with $N=1$ and $\phi=0$, is compared with a random phase field with a number of components $N=1000$. This configuration gives rise to a time-varying amplitude with statistical properties described in previous studies \cite{Zhao2017, Zhou2018, Follett2019, Liu2023}. Two sets of simulations are carried out with a wide range of bandwidths, one with a time-averaged intensity kept to $I_l=1\times10^{15}~\Wcmsqd$ and a second at the lower intensity of $0.5\times10^{15}~\Wcmsqd$.

\begin{table}[!ht]
\begin{tabular}{|l | l|} \hline
 \multicolumn{2}{|c|}{Laser parameters }\\ \hline
 Central laser wavelength & $\lambda_0 = 530~\mathrm{nm}$ \\
 Intensity ramp ($t<t_r$) & $\exp[-9(1 - t/t_r)^2] $\\
 Ramp time & $t_r=1~\mathrm{ps}$\\
 Time-averaged $I_l$ $(t>t_r)$& $I_l=0.5~\&~1.0\times10^{15}~\Wcmsqd$\\
 Laser amplitude & $a_0 = eE_0/m_e\omega_0c=0.010~\&~0.014$\\
 Bandwidth $\Delta\lambda_{\mathrm{b}}$ (nm) & 1, 2, 5, 10, 15, 20, 25, 50 \\
 $N\omega_b/\omega_0$ (\%) & 0.19, 0.38, 0.94, 1.89, \\ & 2.83, 3.78, 4.71, 9.45 \\ \hline
 \multicolumn{2}{|c|}{Plasma parameters} \\  \hline 
 Cell length & $dx=8.4375$~nm, or $\lambda_0/64$ \\ 
 Time step & $dt=0.95~dt_{\mathrm{CFL}}\simeq0.026~\mathrm{fs}$ \\
 Physical box dim. & $L_{\rm sim}=1.66$ mm \\
 Plasma length & $L_p=1.36$ mm \\
 EM B.C. & Silver Muller \\
 Particle B.C. & remove \\
 Nb. particles per cell & 1000 ions, 1000 electrons \\
 Ion species & hydrogen (+1) \\
 Ion temperature & $T_i=200~\mathrm{eV}$\\
 Electron temperature & $T_e=1~\mathrm{keV}$ \\
 Collisions & on for all species \\
 Simulation runtime & $30~\mathrm{ps}$ \\  \hline 
 \end{tabular}
 \caption{Parameters for numerical simulations. The bandwidth $\Delta\lambda_{\rm b}$ is defined as $N\omega_b \lambda_0/\omega_0$.}
\label{tab:simulation_1}
\end{table}

Electron trapping is the main focus of this study. So, we chose a regime of low plasma density $n_e/n_{\rm cr}=0.05$ for a laser with a central wavelength of $\lambda_0= 530~$nm, where the Landau damping of the backward SRS-driven electron plasma wave for $T_e=1$~keV is strong, $\gamma_{\rm pe}/ \omega_{\rm pe} \simeq 0.05$. These parameters correspond to $k_p\simeq 1.7k_0$ and $k_p\lambda_{\rm De} \simeq 0.34$, where $\lambda_{\rm De}$ is the Debye length.

The plasma density has a trapezoidal profile with a plateau of a length $L_p=2572\lambda_0$, which is preceded by a linear density ramp and followed by a ramp down to vacuum over a length of $100 \lambda_0$. The plasma is sandwiched by $150 \lambda_0$ of vacuum at either end of the simulation box. Rather than use an inhomogeneous plasma density~\cite{Guzdar1991} profile, a constant density profile is used so that the resonance length associated with a density scale length is removed as a factor when considering the effect of pump bandwidth on SRS. 

The ion temperature is not of particular importance for this study. It is set to a realistic temperature value of $T_i=200$~eV corresponding to laser interaction with a gas jet. Given the broad range of bandwidths simulated, only a subset of these simulations is shown in figures to aid clarity in the description.

The expected growth rate can be calculated from Eqs. \eqref{eq:gsrs} and \eqref{eq:gdamp} for the case of a monochromatic light. The Landau damping rate $\gamma_{\rm pe}= 37.2~{\rm ps}^{-1}$ is larger than the SRS growth rate without damping, $\gamma_{\rm SRS}= 11.7~{\rm ps}^{-1}$ and the collisional damping rate of plasma waves, which is 0.13~ps$^{-1}$. Consequently, the expected backward SRS growth rate is $\gamma\simeq 3.4~{\rm ps}^{-1}$. According to the analysis in Sec. \ref{sec22}, it can be suppressed with a laser bandwidth of $\Omega_b/2\pi= 8.2$~THz, which corresponds to $\Delta\lambda_{\mathrm{b}}=1.7$~nm.

\subsection{Reflected light}\label{sec32}

\begin{figure}[!ht]
\centering   
\includegraphics[width=\linewidth]{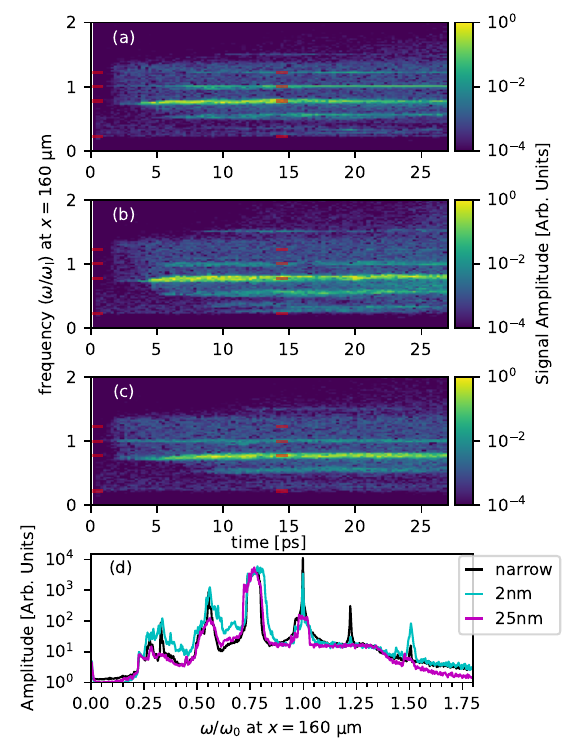}
\caption{Time-resolved spectra of the reflected light normalized to the incident laser amplitude for a narrowband case (a), bandwidth $\Delta\lambda_{\mathrm{b}}=2$~nm (b) 25~nm (c). (d) The time-integrated spectrum of reflected light. The red marks indicate, from highest to lowest frequency: $\omega_0+\omega_{\rm pe}$, $\omega_0$, $\omega_0-\omega_{\rm pe}$, and $\omega_{\rm pe}$. The spectrograms are produced using a Gaussian window of $8000 dt$ with a standard deviation of $2600 dt$ without overlap, the time sampling is performed at $4 dt$. Laser intensity is $1\times10^{15}~\Wcmsqd$.}
\label{fig:ref_spec_var_amp}
\end{figure}

The laser backscattering and transmission are measured in a vacuum close to the entry and exit boundaries, at $x=5dx$ and $x=L_{\rm sim}-5dx$. The measurement in vacuum allows for the separation of forward and backward traveling radiation by computing the Fourier spectra of the fields $E_z+cB_y$ and $E_z-cB_y$, respectively. The time-resolved spectra for the reflected light are shown in \cref{fig:ref_spec_var_amp}. It contains six spectral lines ranging from $\sim 0.25\omega_0$ to $1.5\omega_0$. The two strongest lines at $\omega_0-\omega_{\rm pe}$ and $\omega_0$ correspond to Raman and Brillouin backscattering. The two lines of a smaller intensity at $\omega_0+ \omega_{\rm pe}$ and $\omega_0-2\omega_{\rm pe}$ are related to secondary processes. The secondary forward SRS instability driven by the primary backward propagating scattered wave at the frequency $\omega_0-2\omega_{\rm pe}$ is visible in all simulations, but the intensity of the transmitted signal is an order of magnitude smaller than the primary SRS. The weak line at $\omega_0+ \omega_{\rm pe}$ is seen only in the narrowband case. It is identified as an anti-Stokes component of backward SRS. 

A distinct low frequency reflected signal at $\sim 0.25\omega_0$ can be seen in \cref{fig:ref_spec_var_amp}(d) in the simulation with $\Delta\lambda_{\mathrm{b}} = 2$~nm. It can be related to the linear transformation of the backward propagating plasma wave at the descending plasma density profile. A narrow spectral line at $\sim \omega_0+ 2\omega_{\rm pe} \simeq 1.5\omega_0$ can also be seen in the spectrum. However, the amplitudes of both signals are rather weak.

\begin{figure}[!ht]
 \centering
\includegraphics[width=\linewidth]{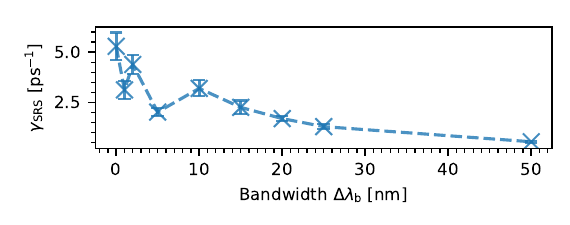}
\caption{Dependence of the growth rate of backward SRS on the laser bandwidth in a plasma with density $n_e/n_{\rm cr}=0.05$ and $T_e=1$~keV. The error bars correspond to fitting errors. The growth rate is calculated from the amplitude of the signal at a frequency $\omega=\omega_0-\omega_{\rm pe}$ within a spectral window of $\pm5$\%. Laser intensity is $1\times10^{15}~\Wcmsqd$.} \label{fig:BSRS_growth_1000eV}
\end{figure}

The spectral line at $\omega_0-\omega_{\rm pe}$ corresponding to backward SRS can be seen in every case, and for the whole time of the simulation. The dependence of the growth rate of backward SRS on the laser bandwidth is shown in \cref{fig:BSRS_growth_1000eV}. The growth rate is calculated by fitting the spectral line with a sinc function in the spectral range of $\pm5$\% and interpolating the temporal evolution of the amplitude with an exponential function.
In the case of a monochromatic pump, it is in good agreement with the theoretical estimate \eqref{eq:gdamp}, but, unexpectedly, it does not notably decrease with the laser bandwidth until $\Delta\lambda_{\mathrm{b}}>15~\mathrm{nm}$ . 

\begin{figure}[!ht]
  \centering
\includegraphics[width=\linewidth]{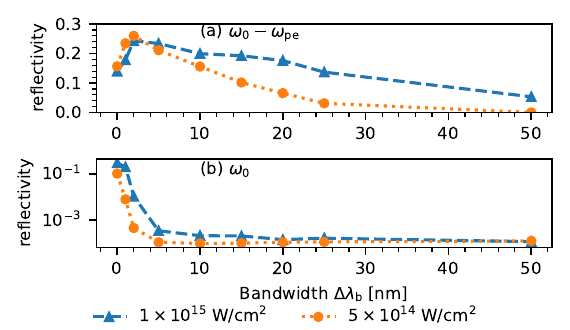}
\caption{Dependence of the reflectivity at frequencies $\omega_0-\omega_{\rm pe}$ (a) and $\omega_0$ (b) on the laser bandwidth. Each spectral component is fitted with a sinc function and the reflectivity is calculated as a ratio of the integral over the line profile to the total laser energy entering the simulation box. Laser intensity is $1\times10^{15}~\Wcmsqd$ (blue) and $0.5\times10^{15}~\Wcmsqd$ (orange).} \label{fig:homogeneous_back_reflection}
\end{figure}

The backward SRS reflectivity in function of laser bandwidth is shown in \cref{fig:homogeneous_back_reflection}(a). It contains two parts: the reflectivity increases with the bandwidth for $\Delta\lambda_{\mathrm{b}}\lesssim 2$~nm and gradually decreases for larger bandwidth. The increase of SRS reflectivity is correlated with a strong reduction of the reflected light at the frequency close to the laser frequency, $\omega_0$. This is explained by a reduction in SBS for a pump bandwidth larger than 5~nm. The reduction in SBS reflectivity can also be seen in the corresponding absence of ion-acoustic waves in simulations with pump bandwidth larger than 5~nm, as discussed in Sec. \ref{sec34}.

The inhibition of the backward SRS reflectivity with a bandwidth increase for $\Delta\lambda_{\mathrm{b}}\gtrsim 2$~nm 
shown in \cref{fig:homogeneous_back_reflection}(a) can be interpreted as a decrease in the spectral pump intensity, which reduces the coupling efficiency to relevant plasma frequencies. This suggestion is confirmed with additional simulations using a lower pump intensity of $I_l=5\times10^{15}~\Wcmsqd$ corresponding to a normalized pump amplitude of $a_0=0.010$. As shown in \cref{fig:homogeneous_back_reflection}(a), the SRS suppression with the bandwidth increase is stronger at a lower laser intensity. While a similar SRS reflectivity is found at the bandwidth of 2~nm, the mitigation is more efficient, and SRS is strongly suppressed at $\Delta\lambda_{\mathrm{b}}=25$~nm. 

The competition between the SBS and SRS response is evident in these simulations. As shown in \cref{fig:homogeneous_back_reflection}, SBS is switched off for $\Delta\lambda_{\mathrm{b}}\gtrsim5$~nm, while the SRS signal peaks up. This can be understood as follows: the SBS instability for a monochromatic light reflects a significant quantity of incident energy and depletes the field that can drive SRS. By contrast, the broadband pump suppresses SBS, thus providing the full pump field for driving SRS. For lower intensity the maximum SBS reflectivity is lower and stronger suppressed at larger bandwidths. The apparent independence of SRS reflectivity on laser intensity at small bandwidths may be explained by the competition with SBS: An increase in laser intensity leads to an increase in SBS reflectivity, so the SRS reflectivity remains on a similar level.

\subsection{Transmission and forward scatter}\label{sec33}
\begin{figure}[!ht]
  \centering
\includegraphics[width=\linewidth]{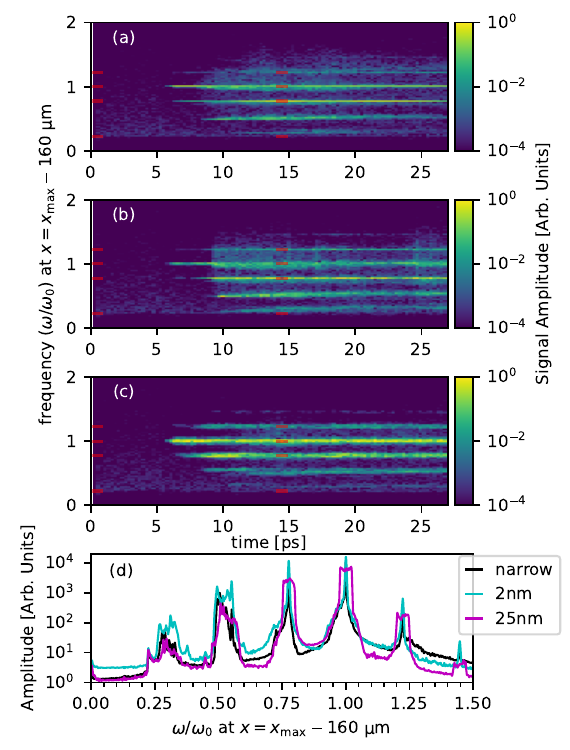}
\caption{Spectra of the transmitted light normalized to the incident laser amplitude for a narrowband case (a), bandwidth of $\Delta \lambda_{\mathrm{b}}=10$~nm (b) and 25~nm (c). (d) The time-integrated spectrum of transmitted light. The red marks indicate, from highest to lowest frequency: $\omega_0+ \omega_{\rm pe}$, $\omega_0$, $\omega_0-\omega_{\rm pe}$, and $\omega_{\rm pe}$. The spectrograms are produced in the same way as \cref{fig:ref_spec_var_amp}. Laser intensity is $1\times10^{15}~\Wcmsqd$.} \label{fig:trans_spec_var_amp}
\end{figure}

\begin{figure}[!ht]
  \centering
\includegraphics[width=\linewidth]{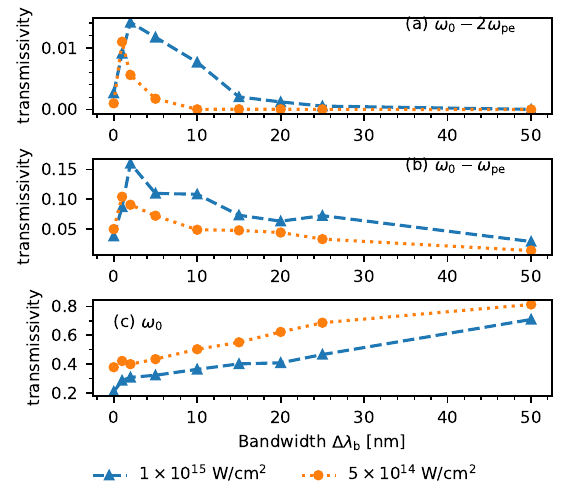}
\caption{Transmissivity at frequencies $\omega_0-2 \omega_{\rm pe}$ (a), $\omega_0- \omega_{\rm pe}$ (b), and $\omega_0$ (c), as a function of bandwidth. Transmission is calculated in the same way as in \cref{fig:homogeneous_back_reflection} by integrating the spectral profiles fitted by a sinc function and dividing them by the total laser energy (the energy transmitted through a vacuum of the same length as simulation box by the end of the simulation). Laser intensity is $1\times10^{15}~\Wcmsqd$ (blue) and $0.5\times10^{15}~\Wcmsqd$ (orange).} \label{fig:homogeneous_transmission}
\end{figure}

The time-resolved spectra for the transmitted light in each simulation can be seen in \cref{fig:trans_spec_var_amp}. In all cases, the signal at the main laser frequency is observed, and its intensity increases with the laser bandwidth. The signal at $\omega_0 - \omega_{\rm pe}$ corresponds to the forward SRS driven by the main laser pulse. The signals at $\omega_0+ \omega_{\rm pe}$ and $\omega_0 -2\omega_{\rm pe}$ are weak and have comparable intensities. Additional low-frequency signal close to $\omega_{\rm pe}$ can be seen in the simulation with $\Delta\lambda_{\mathrm{b}} = 10$~nm. Similar to the reflected signal, it may be related to the linear conversion of the electron plasma wave at the descending part of the plasma density profile.

The time-integrated transmitted laser energy through the plasma is shown in \cref{fig:homogeneous_transmission}. The transmission at $\omega_0$ increases with the bandwidth to a level of $\sim 0.7$ at $\Delta\lambda_{\mathrm{b}}=50$~nm. However, at a lower bandwidth, a significant reduction in transmission at $\omega_0$ is seen, with an increase in transmission at $\omega_0-\omega_{\rm pe}$ to a level of 15\%. This is consistent with an increase in reflectivity at $\omega_0-\omega_{\rm pe}$ seen in \cref{fig:homogeneous_back_reflection}. Therefore, suppression of SBS prompts a stronger backward and forward SRS. At the modest bandwidth of 2\%, the laser energy at the frequency $\omega_0$ is transmitted with an efficiency of $\sim40$\%, the efficiency of backward SRS is $\sim20$\%, and $\sim12$\% is transmitted with forward SRS. The remaining part of laser energy is transferred to plasma electrons. 

As shown in \cref{fig:homogeneous_transmission}, the transmission at frequencies $\omega_0-\omega_{\mathrm{pe}}$ and $\omega_0-2\omega_{\mathrm{pe}}$ has a weaker dependence on bandwidth with a higher intensity pump laser than with the lower intensity case.

\subsection{Plasma wave characterization}\label{sec34}

\begin{figure}[!ht]
  \centering
\includegraphics[width=\linewidth]{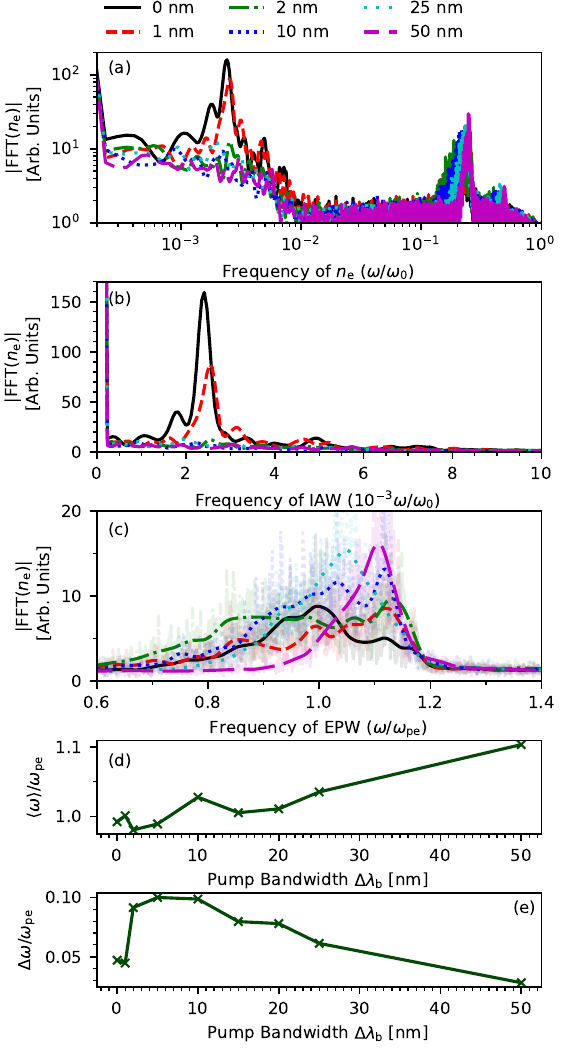}
\caption{(a) Spectrum of the electron density measured in plasma at $x=160~\mu$m from the simulation boundary. The signal is smoothed over a spectral window of $\Delta\omega/\omega_{\rm pe} = 0.23$\% to remove noise with the original signal resolution being 0.026\%. (b) The same spectra as (a) but zoomed to the region $\omega/\omega_0<0.01$ close to the frequency of SBS-excited ion-acoustic wave. (c) The same spectra as (a) but zoomed to frequencies close to the electron plasma frequency. Solid lines show the signal smoothed over a window of 17\%. The signal with less smoothing of 0.23\% is plotted as a transparent line behind. (d) Dependence of the central frequency of plasma waves on the bandwidth found using a Gaussian fit. (e) Dependence of the spectral width of plasma waves found using a Gaussian fit.
} \label{fig:EPW_raw_spectrum}
\end{figure}

In this section, the properties of plasma waves generated in laser-plasma instabilities, SRS and SBS, are discussed. In particular, the plasma waves associated with the SRS backscatter instability are focused on.

The plasma wave spectra shown in \cref{fig:EPW_raw_spectrum} are calculated from the electron density measured as a function of time at a single location close to the entry point of the laser, and so show only plasma waves generated at this point in space. The simulations using the laser bandwidth $\lesssim 1$~nm show distinct low-frequency oscillations with frequency $\sim 0.011\omega_{\rm pe}$, which is absent in simulations with a larger bandwidth. This is the frequency of ion-acoustic wave driven by backward SBS, $2k_0v_{\rm iaw}\simeq 9.05$~ps$^{-1}$, where $v_{\rm iaw} \simeq 3.9\times10^5$~m/s is the ion-acoustic velocity. Since this spectral line disappears in the simulations with a larger bandwidth $\gtrsim2$~nm, it is sufficient to fully suppress SBS in the low-density homogeneous plasmas studied here. The wavelength bandwidth $\Delta\lambda_{\mathrm{b}}=2$~nm corresponds to the frequency bandwidth $N\omega_b \simeq 13$~ps$^{-1}$, which is larger than the SBS growth rate of 2.45~ps$^{-1}$ calculated from the textbook formula \cite{Kruer2003, Tikhonchuk2024}
\begin{equation}\label{eq:gsbs}
\gamma_{\mathrm{SBS}} = \frac{e E_0 k_0\omega_{\rm pi}}{2m_e \omega_0 \sqrt{2\omega_0 k_0v_{\mathrm{Te}}}}.
\end{equation}
That is, the pump bandwidth required to suppress SBS is approximately five times larger than the growth rate, which is larger but compatible with the theoretical formula \eqref{eq8d}.

\begin{figure}[!ht]
  \centering
  \includegraphics[width=\linewidth]{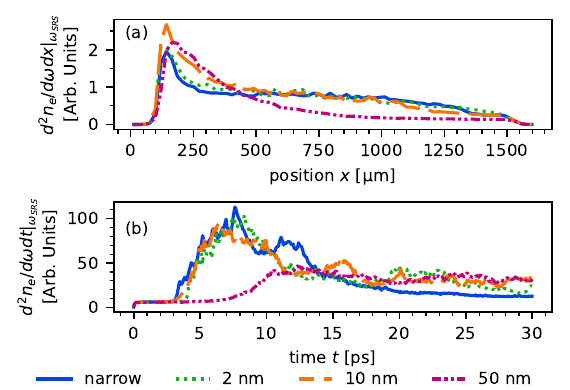}
  \caption{(a) Distribution of the SRS plasma wave amplitude in the plasma. (b) Dependence of the maximum amplitude of the SRS-excited plasma wave on time.}
  \label{fig:SRS_penetration_duration}
\end{figure}

As shown in \cref{fig:EPW_raw_spectrum}(c) and (d), the plasma waves driven by the backward SRS have a broad spectrum shifting to a larger frequency at larger bandwidths. 
This can be interpreted as a negative frequency shift due to the electron trapping occurring at small bandwidths, which is reduced at larger bandwidths. Similarly, the spectral width shown in panel (e) has a maximum of about 10\% at small bandwidth, and gradually decreases to about 3\% at large bandwidth. This can be compared with the laser bandwidth which increases from 1\% at $\Delta\lambda_{\mathrm{b}}=5$~nm to 10\% at $\Delta\lambda_{\mathrm{b}}=50$~nm. 

The plasma waves driven by backward SRS extend far into the plasma. As shown in \cref{fig:SRS_penetration_duration}(a), the plasma wave with $k_p\simeq1.7k_0$ is excited near the laser entry side ($x=150~\mathrm{\mu m}$) and extends up to the far side ($x=1.51~\mathrm{mm}$) in every case except the largest bandwidth. The amplitude of these plasma waves varies over time. As shown in \cref{fig:SRS_penetration_duration}(b), the plasma wave amplitude increases for the first 10 ps, and then slowly decreases. The narrowband simulation shows a distinct second peak, likely as a result of competition with SBS. In contrast to the narrowband case, the simulations with $\Delta\lambda_{\rm b}>2~\mathrm{nm}$ show an almost constant plasma wave amplitude for $t>10$~ps. The initial higher amplitude plasma waves are likely depleted by the trapped electrons they accelerate within the first 10~ps of the simulation (see \cref{fig:SRS_penetration_duration}(b)). The depletion of the SRS plasma wave after 10~ps would need to be balanced by the growth rate of SRS for the plasma wave amplitude to remain constant later in time as is seen in the simulations.

\subsection{Electron heating and trapping}\label{sec35}

\begin{figure}[!ht]
\centering \includegraphics[width=\linewidth]{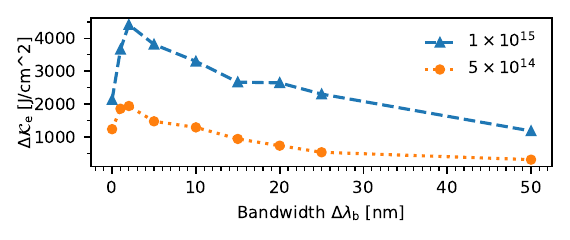}
\caption{Dependence of the total energy gained by electrons at $t=25$~ps on the laser bandwidth. The spatial integration is limited to the constant density plasma plateau and does not include ejected electrons. The total laser energy injected at $t=25$~ps is 25~kJ\:cm$^{-2}$ for $I_{\mathrm{l}}=\times10^{15}\Wcmsqd$. }\label{fig:electron_energy_gain}
\end{figure}

\begin{figure*}[!ht]
  \centering
\includegraphics[width=\textwidth]{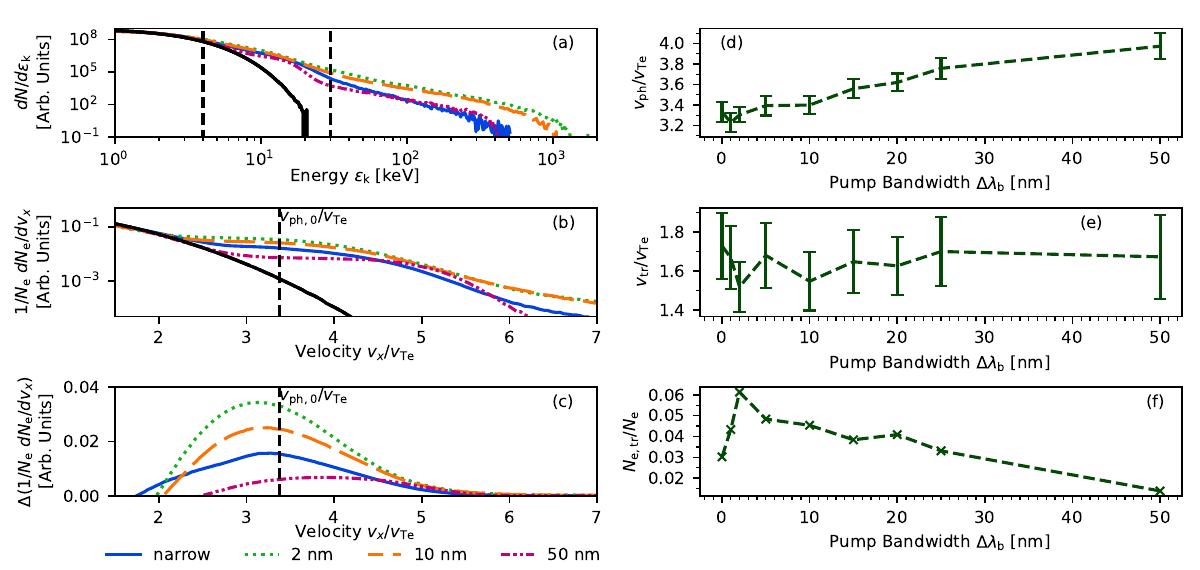}
\caption{(a) Energy distribution of electrons in the constant density region of the plasma taken at $t=25$~ps. The distribution at the initial temperature is shown in black. Vertical dashed lines indicate an interval of 4-30~keV. (b) Electron velocity distribution function close to the phase velocity of the plasma wave driven by backward SRS. Color code is the same as in panel (a). (c) Electron velocity distribution with the initial Maxwellian distribution subtracted. (d) Dependence of the average velocity of trapped electrons $v_{\mathrm{av,tr}}$ on the laser bandwidth calculated from fitting function \eqref{eq:ffit}. (e) Dependence of the trapping velocity $v_{\rm tr}$ on the laser bandwidth. (f) Dependence of fitted value for the total number of trapped electrons on the pump bandwidth.} 
\label{fig:edist_vel} 
\end{figure*}

The energy transferred from laser to electrons, $\Delta\mathcal{K}_e$, is calculated by integrating over the electron energy distribution at a time $t$ and subtracting a similar value at $t=0$. The integration is limited to the constant density plateau, and so neglects the small quantity of ejected electrons. As shown in \cref{fig:electron_energy_gain}, the bandwidth dependence of electron energy correlates well with the SRS reflectivity in \cref{fig:homogeneous_back_reflection}(a). The electron energy gain, $\Delta\mathcal{K}_e \sim 2$~kJ\:cm$^{-2}$ at $t=25$~ps corresponds to approximately 8\% of the total laser energy of 25~kJ\:cm$^{-2}$ (for $I_{\mathrm{l}}=\times10^{15}\Wcmsqd$), in the narrowband case. It increases to 20\% in the case of a 2~nm bandwidth laser and then slowly decreases with increasing bandwidth, only reaching the same level as the narrowband case at a bandwidth of 25~nm. While the laser does not fully propagate through the plasma at $t=25$~ps, the energy transferred to electrons is comparable to the transmitted laser energy, given the time taken to traverse the simulation box through the plasma and the 1~ps laser rise time. The peak in electron energy gain at $\Delta\lambda_{\mathrm{b}} =2$~nm corresponds to the full SBS suppression, and also to the rise in backward SRS scattering as a function of bandwidth seen in \cref{fig:homogeneous_back_reflection}. Therefore, the electron energy gain can be related to their acceleration and heating in the SRS-driven plasma wave. 

The electron energy distribution is shown in \cref{fig:edist_vel}(a). Compared to the initial Maxwellian distribution, one can distinguish two groups of suprathermal electrons. The energy interval delimited by dashed vertical lines corresponds to the electrons that directly interact with the backward SRS-driven plasma wave. The electrons with larger energies are accelerated in the secondary plasma waves or plasma waves driven by the forward SRS. The suppression of SBS is manifested in the electron energy spectrum by an increase of the number of accelerated electrons with energies $\varepsilon_e>30$~keV.

The electron trapping in the SRS-driven plasma wave is manifested in \cref{fig:EPW_raw_spectrum}(d) and (e) with a notable shift and broadening of the plasma wave spectra, which are spread over a wide frequency range from 0.8 to $1.2\omega_{\rm pe}$. Another signature of electron trapping can be seen in the plateau-like formation in the electron energy distribution shown in \cref{fig:edist_vel}(a) in the range of $\sim4-30$~keV. 

\begin{figure}[!ht]
  \centering
\includegraphics[width=0.95\linewidth]{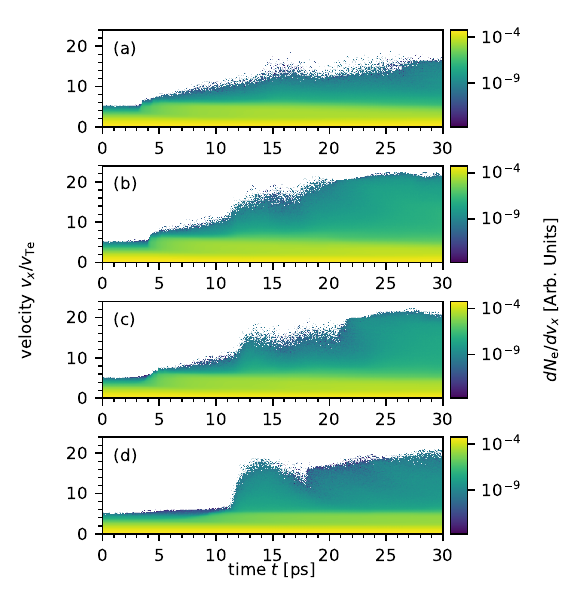}
\caption{Time-resolved electron velocity distributions for simulations with a narrowband (a) and broadband lasers with $\Delta\lambda_{\mathrm{b}}=2$~nm (b), 25~nm (c), and 50~nm (d). The spectra are calculated over the constant density plasma region. \label{fig:elec_dist_var_amp_1kev} }
\end{figure}

A quantitative analysis of electron trapping is demonstrated in \cref{fig:edist_vel}(b) and (c). A plateau in the electron distribution shown in panel (b) is formed around the phase velocity of the plasma wave driven by backward SRS. The electron velocity distribution is plotted on a linear scale on panel (c) with the initial distribution subtracted. A peak of non-thermal electrons can be seen close to the plasma wave phase velocity indicated with a vertical dashed line. The width of the trapping region and the central velocity are evaluated by using a fitting distribution, 
\begin{equation}\label{eq:ffit}
 f(v)=\mathrm{sinc}(\pi (v-v_{\rm av,tr})/v_{\rm tr}). 
\end{equation} 
The dependence of the average velocity on the pump bandwidth shown in \cref{fig:edist_vel}(d) can be interpreted as a nonlinear shift of the plasma wave frequency, which is shown in \cref{fig:EPW_raw_spectrum}(d). It increases from $3.4v_{\rm Te}$ to $\sim 4v_{\rm Te}$ with the laser bandwidth increasing up to 50~nm. The plateau width obtained from the fitting function is shown in \cref{fig:edist_vel}(e). It is approximately constant across the whole range of laser pump bandwidths tested, with $v_{\rm tr}\simeq 1.6 v_{\rm Te}$. 

This trapping velocity can be associated with the amplitude of the plasma wave $E_x$ as \cite{Berger2013, Tikhonchuk2024}:
\begin{equation}\label{eq:Etrap}
 m_ev_{\rm tr}^2/2 \simeq 2eE_x/k_p. 
\end{equation} 
According to this relation, the dimensionless amplitude of the plasma wave, $eE_x/m_e\omega_{\rm pe} v_{\rm Te}$ slightly decreases from 0.19 to 0.16 when the laser bandwidth increases from 5~nm to 50~nm. This is compatible with the decrease of SRS reflectivity, \cref{fig:homogeneous_back_reflection}(a), and electron kinetic energy, \cref{fig:electron_energy_gain}, with the increase of laser bandwidth. These correlations confirm that the backward SRS is the origin of electron heating and acceleration. The fact that the phase velocity increases with increasing laser bandwidth explains the reduction of energy gain by electrons, as there are fewer electrons at a higher phase velocity. Additionally, at higher bandwidths, fewer electrons are trapped within the plasma waves (see \cref{fig:edist_vel}(f)), which is related to the higher phase velocity at higher bandwidths.

The time-resolved electron velocity distribution is shown in \cref{fig:elec_dist_var_amp_1kev}. A small plateau is formed from $2v_{\rm Te}$ to $\sim 6v_{\rm Te}$ at a time between 4~ps for the narrowband laser to 10~ps for the bandwidth of $\Delta\lambda_{\mathrm{b}}=10$~nm, indicating that the electron trapping is delayed for wider bandwidth. Moreover, electrons with high velocities from $10v_{\rm Te}$ to $20v_{\rm Te}$ are produced at a time of 10~ps. However, the number and energy of this electron population are small, about 3 orders of magnitude smaller than that of the trapped electron population.

\begin{figure}[!ht]
  \centering
\includegraphics[width=\linewidth]{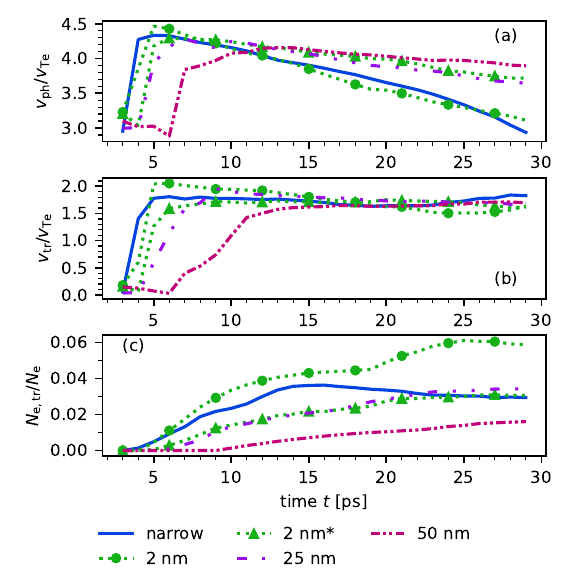}
\caption{Time dependence of the average velocity of trapped electrons (a), width of the trapping zone (b), and fraction of trapped electrons (c) for the simulations with different laser bandwidths (dashed color lines). Solid blue line shows the same quantities obtained in the narrow band simulation. Laser intensity is $10^{15}~\Wcmsqd$ except the case of $\Delta\lambda_{\mathrm{b}}=2$~nm marked with $\ast$ and green triangles calculated for $I_l=5\times10^{14}~\Wcmsqd$.} \label{fig:pwave_v_time}
\end{figure}

Temporal evolution of the trapped electron characteristics is shown in \cref{fig:pwave_v_time} for different laser bandwidths. After the initial sharp increase, the average \sout{ phase} velocity, $v_{\mathrm{av,tr}}$ slowly decreases with time, while the width of the trapping zone, $v_{\mathrm{tr}}$, remains constant. The number of trapped particles, $n_{\mathrm{tr}}$, decreases as the bandwidth increases. There is a correlation in the temporal evolution of all three quantities: an increase in laser bandwidth above 10~nm leads to a time delay and a slower growth in the number of trapped particles, accompanied by a slower decrease of the average velocity of trapped electrons. As the number of trapped electrons continuously increases and their average velocity continuously decreases up to the termination of the simulation, the SRS saturation is not achieved in the simulations with $\Delta\lambda_{\mathrm{b}}=25$ and 50~nm. It seems possible that the number of trapped electrons might attain similar values independent of the pump bandwidth.

The effect of bandwidth on the growth rate of backward SRS is independent of the laser amplitude in the regime of exponential growth, and the saturation of SRS is not expected to be a function of laser amplitude either. The reduction of the effect of laser bandwidth on SRS saturation at $\Delta\lambda_{\mathrm{b}}\gtrsim 2$~nm may be partly explained by the increase in electron trapping seen in the higher amplitude pump simulations. A comparison of the three trapping parameters, $ v_{\mathrm{ph}}$, $v_{\mathrm{tr}}$, and $n_{\mathrm{tr}}$, can be seen in \cref{fig:pwave_v_time} for the bandwidth of 2~nm and intensities $I_l=10^{15}~\Wcmsqd$ and $5\times10^{14}~\Wcmsqd$. While the trapping velocity remains unchanged, the number of trapped electrons decreases and their average velocity increases with the laser intensity.

There is a correlation between the bandwidth dependence of the average frequency of the SRS-driven plasma wave, \cref{fig:EPW_raw_spectrum}(d), and the average velocity of trapped electrons, \cref{fig:edist_vel}(d). 
The rise in frequency can be interpreted as a restoration of the plasma wave back to that predicted by the linear Bohm-Gross dispersion as the trapped electron population decreases.

The simulations with bandwidth $2-25$~nm and intensity $I_l=10^{15}~\Wcmsqd$ show the largest number of trapped electrons and the largest reduction of the plasma wave phase velocity. This is correlated with the largest plasma wave spectral width, and lowest frequency as shown in \cref{fig:EPW_raw_spectrum}. Crucially, these simulations also show the lowest sensitivity of SRS reflectivity to bandwidth (see \cref{fig:homogeneous_back_reflection}). Therefore, electron trapping appears to be the main process responsible for the decrease in sensitivity of SRS reflectivity to laser pump bandwidth. This is related to the frequency broadening of the plasma wave (see \cref{fig:EPW_raw_spectrum}) which facilitates the frequency and wavenumber matching conditions required for SRS excitation.

\subsection{Secondary instabilities and forward SRS}\label{sec36}

The generation of high-energy electrons is likely due to a synergy of several laser-plasma instabilities. Firstly, electrons are trapped in the plasma wave produced by backward SRS, and then some of them are accelerated in the interaction with the plasma wave driven by forward SRS and/or secondary backward SRS. These processes are activated when SBS is suppressed (as can be demonstrated in the transmitted, \cref{fig:trans_spec_var_amp}, and reflected light, \cref{fig:ref_spec_var_amp}). 

\begin{figure}[!ht]
  \centering
\includegraphics[width=0.48\textwidth]{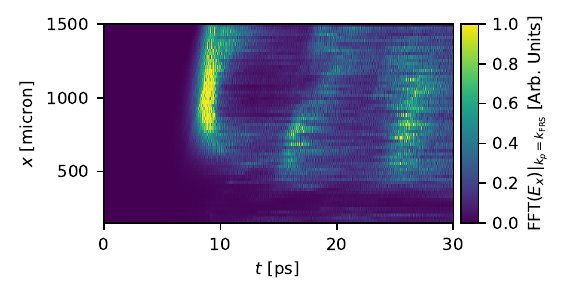}
\caption{Spatial and temporal evolution of plasma wave electric field $E_x$ with wavenumber $k=0.23\omega_0/c$. A spectrogram of $E_x$ is obtained with a Gaussian window of $6000~dx$ and dispersion of $1000~dx$. Laser intensity is $10^{15}~\Wcmsqd$ and bandwidth is 25~nm.} \label{fig:FRS_temp_1000eV_25nm}
\end{figure}

The electric field of the plasma wave driven by forward SRS corresponding to the plasma wave number $k_\mathrm{FSRS}= 0.23\omega_0/c$ is shown in \cref{fig:FRS_temp_1000eV_25nm}. This plasma wave grows in a region relatively far away from the laser entry boundary at a time when the backward SRS has already saturated. It decays periodically with timing similar to that of the burst of energetic electrons shown in \cref{fig:elec_dist_var_amp_1kev}. The phase velocity of this plasma wave driven by forward SRS, $v_{\rm ph}=0.96c$, is close to the light velocity, so, it can accelerate electrons to relativistic energies. However, this phase velocity is well outside the main electron distribution, and only a very small number of electrons could be excited by this wave.

The phase velocity of the forward SRS plasma wave places it well within the weakly damping regime for a thermal plasma with $T_e=1$~keV, with $\lambda_{\rm De}=16.7$~nm giving $k_\mathrm{FSRS} \lambda_\mathrm{De}=0.045$. The exchange of energy between forward SRS-driven plasma wave and electrons is one way for energy to be drained from the plasma wave until it dissipates and is then excited again by the laser. This dissipation growth cycle can be seen in \cref{fig:FRS_temp_1000eV_25nm} with a period of roughly $5-7$~ps.

\section{Conclusions}\label{sec4}
This study demonstrates the insensitivity of backward SRS to the laser bandwidth in the electron trapping regime. A homogeneous plasma is chosen so that any relationship between resonance length and bandwidth is ignored. Electron trapping in one-dimensional PIC simulations is ubiquitous. Once an electron is caught in a plasma wave potential, it does not have the degrees of freedom required to exit the plasma wave, so the trapping regime is exaggerated in one-dimensional simulations. This limitation has been used as a feature in this study as it allows for the exploration of the physics of broad bandwidth lasers with trapped electrons in a long length ($>1$~mm), long timescale (10s of ps) plasmas. The one-dimensional approach also allows for a more direct demonstration of the specific physical effect of plasma wave properties on the backward SRS instability without additional two-dimensional effects that would otherwise obscure the results.

Stimulated Raman scattering is a three-wave coupling process that requires frequency and wave-number matching of incident and scattered light with plasma waves. Under conditions where the frequency of each of these components is either time-dependent or broadened, the growth of SRS changes. In the kinetic inflation regime, where Landau damping is initially strong, the SRS growth is delayed, thus favoring the SBS excitation, which dominates the plasma reflectivity and reduces the laser wave transmission. Nevertheless, SRS develops, leading to the trapping and acceleration of electrons and the broadening of the plasma wave spectrum. 

The broadening of the laser frequency spectrum modifies this scenario of SRS-SBS competition in a low-density hot plasma. At laser bandwidth exceeding $1-2$~nm, the SBS is fully suppressed, thus favoring a stronger SRS response, manifested in an increase in reflectivity and the number and energy of suprathermal electrons. By contrast, the laser bandwidth affects the SRS excitation much less. Depending on the laser intensity, the bandwidth needed for mitigation of SRS is in the range of $20-50$~nm. Such a weak sensitivity of SRS on the laser bandwidth in the inflation regime is explained by a strong modification of the electron distribution function in the resonance zone which is accompanied by the broadening of the spectrum of plasma waves.

\section{Acknowledgments}
Numerical calculations using the code SMILEI were performed in part on the Karolina cluster managed by IT4Innovations. This work was supported by the Ministry of Education, Youth and Sports of the Czech Republic through the e-INFRA CZ (ID:90254).
\bibliography{broadband}

\end{document}